\documentclass[10pt,a4paper]{article}

\usepackage[T1]{fontenc}
\usepackage[utf8]{inputenc}
\usepackage{mathptmx}          
\usepackage{times}             
\usepackage[
  a4paper,
  top=2.0cm,
  bottom=2.0cm,
  left=2.4cm,
  right=2.4cm,
  headsep=0.8cm,
  footskip=1.2cm
]{geometry}

\usepackage{amsmath,amssymb,amsfonts}
\usepackage{bm}                

\usepackage{graphicx}
\usepackage{float}

\usepackage{booktabs}
\usepackage{array}
\usepackage{multirow}

\usepackage[
  colorlinks=true,
  linkcolor=black,
  citecolor=green,
  urlcolor=blue
]{hyperref}

\usepackage{setspace}
\usepackage{ragged2e}          
\usepackage{microtype}         
\usepackage{changepage}        
\usepackage{caption}
\usepackage{subcaption}
\usepackage[numbers]{natbib}   
\usepackage{xcolor}            

\usepackage{titlesec}

\titleformat{\section}[block]
  {\normalfont\normalsize\centering}
  {\thesection.}
  {1em}                        
  {\MakeUppercase}

\titleformat{\subsection}[block]
  {\normalfont\itshape\normalsize\centering}
  {}
  {0em}
  {}

\titlespacing{\section}{0pt}{14pt}{6pt}
\titlespacing{\subsection}{0pt}{10pt}{4pt}

\newenvironment{docabstract}{%
  \list{}{\leftmargin=0.4cm\rightmargin=0cm}%
  \item\relax
  \textbf{Abstract}---\ignorespaces
}{%
  \endlist\vspace{2pt}%
}

\newcommand{\keywords}[1]{%
  \vspace{4pt}
  \hspace{0cm}\textbf{Keywords:} #1\par
  \vspace{10pt}%
}
\newenvironment{bodycontent}{%
  \begin{adjustwidth}{0.6cm}{0.6cm}%
}{%
  \end{adjustwidth}%
}

\begin{document}

\begin{center}
  {\fontsize{14}{17}\selectfont\bfseries
   Towards Robust Arabic Speech Emotion Recognition with Deep Learning
  }
\end{center}

\vspace{8pt}
\begin{center}
  {\normalsize
    Youcef S. Gheffari$^{a}$,\quad
    Samiya Silarbi$^{a}$
  }
\end{center}

\vspace{6pt}
\begin{center}
  {\small\itshape
    $^{a}$ADASCA Laboratory – Advanced Data Science and Cognitive Applications,\\
    Université des Sciences et de la Technologie d'Oran Mohamed Boudiaf (USTO-MB),
    Oran, Algeria\\[1pt]
    *e-mail: \href{mailto:samiya.silarbi@univ-usto.dz}{samiya.silarbi@univ-usto.dz}%
  }
\end{center}

\footnotetext[1]{Submitted to \textit{Mathematical Models and Computer Simulations}.}

\vspace{10pt}

\begin{bodycontent}

\begin{docabstract}
Speech Emotion Recognition (SER) aims to identify a speaker’s emotional state from audio signals. While recent advances in deep learning have significantly improved SER performance in Indo-European languages, Arabic SER remains underexplored and challenging due to dialectal diversity, limited annotated datasets, and the difficulty of modeling both local spectral cues and long-range temporal dependencies.

To address these limitations, this study investigates whether hybrid architectures that jointly model spatial and contextual information can improve emotion recognition in Arabic speech. We propose and evaluate a comparative framework involving three architectures: a CNN-LSTM model, a CNN-Transformer model, and a fine-tuned wav2vec 2.0 model. The first two models leverage MFCC and spectrogram-based representations, while wav2vec 2.0 operates directly on raw audio through self-supervised representations.

Experiments conducted on the EYASE and BAVED datasets demonstrate that the proposed CNN-Transformer architecture significantly outperforms the other models, achieving an accuracy of 98.1\%. This result highlights the effectiveness of combining convolutional feature extraction with Transformer-based global context modeling.

The main contribution of this work lies in providing a systematic comparison of hybrid and self-supervised approaches for Arabic SER, and in demonstrating that CNN-Transformer architectures offer a robust solution for capturing both spectral and long-range dependencies in low-resource and dialectally diverse settings.
\end{docabstract}

\vspace{4pt}


\vspace{10pt}

\section{INTRODUCTION}

Understanding human emotions from speech has become a central topic in human–computer interaction, with applications in virtual assistants, mental health monitoring, call-center analytics, and intelligent tutoring systems. Speech Emotion Recognition (SER) focuses on automatically identifying a speaker’s emotional state from acoustic signals by analyzing prosodic and spectral characteristics such as pitch, intensity, and temporal dynamics \cite{ElAyadi2011}. Recent advances in deep learning have substantially improved recognition accuracy; however, most existing approaches are primarily designed and evaluated on Indo-European languages, especially English, leaving languages such as Arabic comparatively underrepresented \cite{Tajalsir2023}.

Arabic poses additional challenges for SER due to its morphological richness, phonetic variability, and the presence of numerous regional dialects. These linguistic properties, combined with the limited availability of large-scale annotated emotional speech corpora, complicate model generalization and robustness \cite{AbdelHamid2020}. Traditional SER systems often rely on hand-crafted acoustic features such as MFCCs and prosodic descriptors combined with classifiers like Support Vector Machines (SVM), which limits their ability to capture complex emotional patterns. More recent approaches include Convolutional Neural Networks (CNN) applied to log-Mel spectrograms, Recurrent Neural Networks such as BiLSTM and CRNN architectures for temporal modeling, and Transformer-based models that leverage attention mechanisms for global context understanding. In addition, self-supervised models such as wav2vec 2.0 and HuBERT have shown strong performance by learning contextualized speech representations directly from raw audio \cite{Ioffe2015}.

Despite these advances, existing Arabic SER studies remain limited in terms of dataset diversity, architectural comparison, and rigorous evaluation under a unified experimental protocol. Most works focus on a single model or dataset, making it difficult to draw reliable conclusions about the relative effectiveness of different architectures, especially in low-resource and dialectally diverse settings \cite{Akccay2020,ElAyadi2011}.

To address this gap, this work investigates and compares three complementary deep learning approaches: CNN–LSTM, CNN–Transformer, and wav2vec 2.0 \cite{Baevski2020}. These models are selected because they represent three distinct paradigms in SER: hybrid spatial–temporal modeling (CNN–LSTM), hybrid spatial–contextual modeling using attention (CNN–Transformer), and end-to-end self-supervised representation learning (wav2vec 2.0). In particular, the CNN–Transformer architecture is of special interest, as it combines the ability of CNNs to extract robust local spectral features with the Transformer’s capability to model long-range dependencies, which is crucial for capturing subtle emotional dynamics in speech \cite{Srivastava2014}.

The comparison of these models within a unified framework is motivated by the need to evaluate how different modeling strategies feature-based, hybrid, and self-supervised perform under the same conditions, using consistent preprocessing, feature extraction, and evaluation protocols. This allows for a fair and systematic assessment of their strengths and limitations in Arabic SER.

The main contributions of this work can be summarized as follows:
\begin{enumerate}
    \item We provide a unified experimental framework for Arabic Speech Emotion Recognition, ensuring consistent preprocessing, feature extraction, and evaluation across models.
    \item We conduct a comprehensive comparison of three representative architectures: CNN--LSTM, CNN--\allowbreak Transformer, and wav2vec~2.0, covering hybrid and self-supervised approaches.
    \item We demonstrate the effectiveness of CNN–Transformer models in capturing both local spectral features and long-range temporal dependencies, achieving superior performance on Arabic datasets.
    \item We provide empirical insights into the suitability of different deep learning paradigms for low-resource and dialectally diverse Arabic speech settings.
\end{enumerate}

The remainder of this paper is organized as follows. Section 2 reviews the related work in Speech Emotion Recognition, with a particular focus on Arabic SER. Section 3 presents the theoretical foundations, including feature extraction techniques and deep learning models. Section 4 describes the proposed architectures, datasets, and experimental setup. Section 5 presents and discusses the experimental results. Finally, Section 6 concludes the study and outlines perspectives for future research in Arabic speech emotion recognition.

\section{Related Work}

Speech Emotion Recognition (SER) has been widely studied using various machine learning and deep learning approaches. In the context of Arabic speech, research remains relatively limited due to challenges such as dialectal diversity and the scarcity of annotated datasets \cite{Meftah2021}. This section reviews existing work by categorizing it into three main groups: traditional approaches, deep learning-based methods, and recent contextual/self-supervised models \cite{Picard2000, DMello2012}.

Early studies in Arabic SER primarily relied on handcrafted acoustic features such as Mel-Frequency Cepstral Coefficients (MFCC), pitch, intensity, and other prosodic descriptors. These features were typically combined with classical machine learning classifiers such as Support Vector Machines (SVM), K-Nearest Neighbors (KNN), Random Forest (RF), and Multilayer Perceptrons (MLP).

For instance, Abdel-Hamid et al. \cite{AbdelHamid2020} introduced the EYASE dataset and applied SVM and KNN classifiers using spectral and prosodic features, achieving moderate accuracy around 64\%. Similarly, Dalal and Kedidi \cite{Dalal2020} used feature selection techniques based on rough set theory combined with MLP classifiers, achieving up to 87\% accuracy.

With the rise of deep learning, Arabic SER has shifted toward models capable of automatic feature extraction. Convolutional Neural Networks (CNN) have been widely used to process spectrogram and log-Mel representations, enabling the extraction of spatial features from speech signals \cite{Tajalsir2023, Mahmoudi2023, Shahin2023, ElSeknedy2023}.

Rakan et al. \cite{Rakan2023} applied CNNs on log-Mel spectrograms and achieved 88.6\% accuracy on the EYASE dataset. Recurrent models such as Long Short-Term Memory (LSTM) and Bidirectional LSTM (BiLSTM) were introduced to capture temporal dependencies in speech signals. Hybrid architectures, including CNN-LSTM and Convolutional Recurrent Neural Networks (CRNN), have shown improved performance by combining spatial and temporal modeling capabilities.

For example, Atila and Şengür \cite{Atila2021} proposed a CNN-LSTM model with attention mechanisms, achieving high accuracy on benchmark datasets. These approaches significantly outperform traditional methods but often require large datasets and careful tuning.

Recent advances in SER have focused on Transformer-based architectures and self-supervised learning models. Transformers leverage attention mechanisms to capture long-range dependencies and contextual relationships in speech signals.

Al-onazi et al. \cite{Alonazi2022} demonstrated the effectiveness of Transformer-based models on the BAVED dataset, achieving high accuracy. In parallel, self-supervised models such as wav2vec 2.0 and HuBERT have emerged as powerful alternatives, learning rich representations directly from raw audio without requiring extensive labeled data.

Ali Mohamed and Salah Aly \cite{Mohamed2021} showed that wav2vec 2.0 improves performance in Arabic SER tasks, achieving an F1-score of around 80\%. These models are particularly promising for low-resource languages like Arabic, as they reduce dependence on large annotated datasets while capturing both local and global speech patterns.

Table~\ref{tab:related_work} summarizes key studies in Arabic SER, highlighting datasets, features, models, and limitations.

\begin{table}[H]
\centering
\caption{Summary of Related Work in Arabic Speech Emotion Recognition}
\label{tab:related_work}
\renewcommand{\arraystretch}{1.3}
\setlength{\tabcolsep}{3pt}

\begin{tabular}{
>{\raggedright\arraybackslash}p{2.7cm}
>{\centering\arraybackslash}p{1.4cm}
>{\centering\arraybackslash}p{1.2cm}
>{\raggedright\arraybackslash}p{2.4cm}
>{\raggedright\arraybackslash}p{2.3cm}
>{\centering\arraybackslash}p{1.9cm}
>{\raggedright\arraybackslash}p{2.9cm}
}

\toprule
\textbf{Article} & \textbf{Dataset} & \textbf{Classes} & \textbf{Features} & \textbf{Model} & \textbf{Metrics} & \textbf{Main Limitation} \\
\midrule

Abdel-Hamid et al. (2020) \cite{AbdelHamid2020}
& EYASE & 4 
& MFCC, prosodic 
& SVM, KNN 
& Accuracy: 64\% 
& Limited performance, shallow models \\
\midrule
Dalal \& Kedidi (2020) \cite{Dalal2020}
& ANAD & 4 
& LLD, MFCC 
& MLP 
& Accuracy: 87\% 
& Small dataset, limited generalization \\
\midrule
Rakan et al. (2023) \cite{Rakan2023}
& EYASE & 4 
& Log-Mel spectrogram 
& CNN 
& Accuracy: 88.6\% 
& Limited temporal modeling \\
\midrule
Atila \& Şengür (2021) \cite{Atila2021}
& RAVDESS & 8 
& Spectrogram 
& CNN-LSTM 
& Accuracy: 89.75\% 
& Tested on non-Arabic data \\
\midrule
Al-onazi et al. (2022) \cite{Alonazi2022}
& BAVED & 5 
& Augmented features 
& Transformer 
& Accuracy: 95.2\% 
& High computational cost \\
\midrule
Mohamed \& Aly (2021) \cite{Mohamed2021}
& BAVED & 5 
& Raw audio 
& wav2vec 2.0 
& F1-score: 80\% 
& Requires fine-tuning resources \\

\bottomrule
\end{tabular}
\end{table}

Despite the progress achieved in Arabic SER, several limitations remain. Existing studies often focus on a single architecture or dataset, making comparisons difficult. Moreover, there is a lack of unified evaluation frameworks that ensure fair comparison across models. In addition, few works systematically evaluate hybrid architectures against self-supervised approaches under the same experimental conditions.

This gap motivates the present work, which aims to provide a unified and systematic comparison of CNN-LSTM, CNN-Transformer, and wav2vec 2.0 models for Arabic speech emotion recognition.

\section{Methodology}

This section briefly presents the essential concepts underlying the proposed Arabic Speech Emotion Recognition (SER) framework. Rather than providing a detailed theoretical review, we focus on the key elements required to justify the architectural choices and experimental design.

\subsection{Problem Formulation}

Speech Emotion Recognition (SER) is formulated as a supervised
multi-class classification problem, where the objective is to infer
the emotional state of a speaker from an input speech signal.

Let $\mathcal{D} = \{(x_i, y_i)\}_{i=1}^{N}$ denote a dataset of $N$
utterances, where $x_i$ represents a speech signal and
$y_i \in \mathcal{Y}$ is its corresponding emotion label drawn from a
set of $C$ predefined classes, $\mathcal{Y} = \{y_1, \dots, y_C\}$.

Each input signal $x_i$ is transformed into a time--frequency
representation using Mel-spectrogram extraction:
\[
X_i \in \mathbb{R}^{F \times T},
\]
where $F$ denotes the number of Mel frequency bins and $T$ represents
the number of temporal frames. For self-supervised approaches, the raw
waveform is used directly, $x_i \in \mathbb{R}^{L}$, where $L$ is the
signal length.

A parametric function $\phi(\cdot; \theta)$ is used to map the input
signal to a latent representation:
\[
h_i = \phi(x_i; \theta), \quad h_i \in \mathbb{R}^{d},
\]

\begin{figure}[H]
\centering
\includegraphics[width=0.85\linewidth]{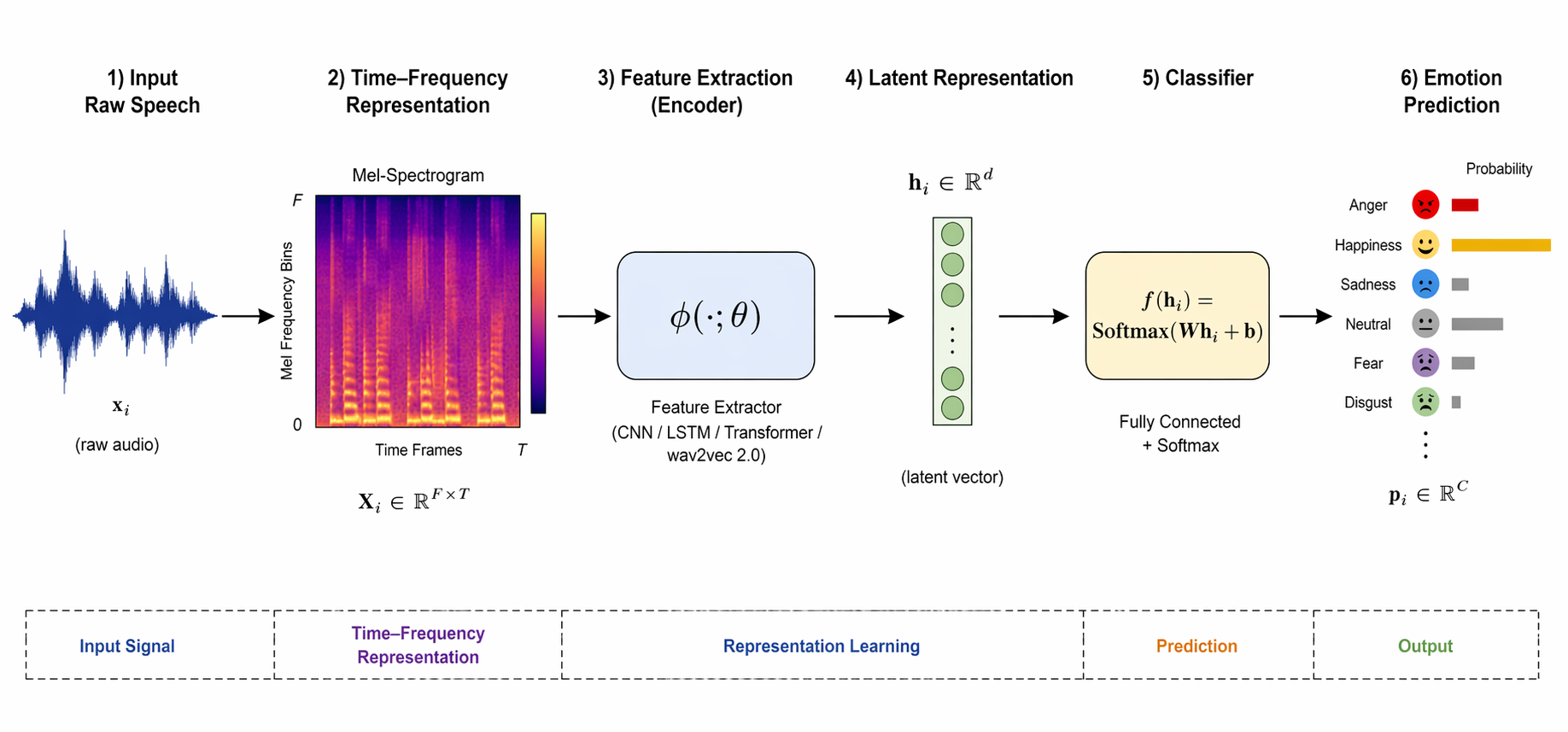}
\caption{Overview of the Speech Emotion Recognition (SER) pipeline. 
The input speech signal is transformed into a time--frequency representation, 
encoded into a latent feature space, and finally mapped to emotion classes.}
\label{fig:problem_formulation}
\end{figure}

where $\theta$ denotes the learnable parameters. This representation
is expected to capture both local spectro-temporal patterns and
long-range temporal dependencies that characterize emotional speech.

The extracted features are then passed to a classifier that produces
a probability distribution over emotion classes:
\[
p_i = \text{Softmax}(W h_i + b),
\]
where $W \in \mathbb{R}^{C \times d}$ and $b \in \mathbb{R}^{C}$ are
trainable parameters. The predicted label is obtained as:
\[
\hat{y}_i = \arg\max_{c \in \mathcal{Y}} p_i^{(c)}.
\]

The model parameters are optimized by minimizing the cross-entropy
loss over the dataset:
\[
\mathcal{L} = - \frac{1}{N} \sum_{i=1}^{N}
\sum_{c=1}^{C} y_i^{(c)} \log p_i^{(c)},
\]
where $y_i^{(c)}$ denotes the one-hot encoding of the ground-truth
label.

Within this formulation, the central challenge lies in learning a
mapping that effectively integrates complementary information across
multiple temporal scales, ranging from short-term spectral cues to
long-range contextual dependencies. To address this, the present work
investigates hybrid and self-supervised architectures, including
CNN--BiLSTM, CNN--Transformer, and wav2vec~2.0, under a unified
experimental framework.

\subsection{Compared Models and Baselines}

To provide a comprehensive and fair evaluation of the proposed
approaches, a set of baseline and comparative models is considered.
These models are organized into four categories, ranging from
traditional machine learning methods to advanced deep learning and
self-supervised architectures.

\paragraph{Traditional Baseline}\mbox{}\\

\textbf{SVM + MFCC} \quad baseline represents a traditional
approach based on handcrafted features. While MFCCs capture general
spectral properties, they lack the capacity to model complex temporal
and contextual dependencies.

Moreover, the fixed feature representation limits the model’s adaptability to diverse emotional patterns. This baseline serves
as a reference point to quantify the advantage of deep learning based approaches.

\begin{figure}[H]
\centering
\includegraphics[width=0.7\linewidth]{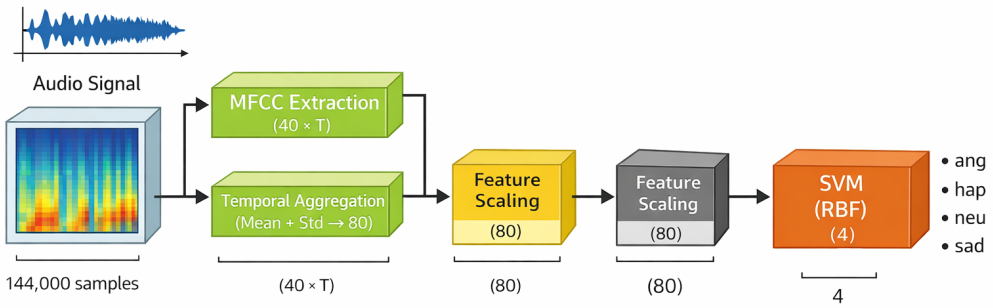}
\caption{SVM + MFCC pipeline using handcrafted features.}
\label{fig:svm_mfcc}
\end{figure}

\paragraph{Simple Deep Learning Baselines}\mbox{}\\

\textbf{CNN}  \quad The model focuses exclusively on local spectro-temporal
patterns. While it is effective at capturing short-term acoustic
structures such as formants and energy variations, it lacks the ability to model long-range temporal dependencies \cite{Krizhevsky2012}. 

As a result, emotional cues that evolve over time, such as gradual pitch changes or prosodic patterns, cannot be fully captured. This baseline therefore highlights the limitations of relying solely on local feature extraction.

\begin{figure}[H]
\centering
\includegraphics[width=0.7\linewidth]{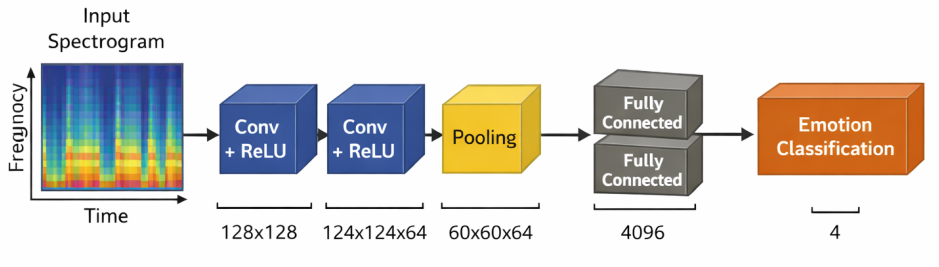}
\caption{CNN-only architecture for local spectro-temporal feature extraction.}
\label{fig:cnn_only}
\end{figure}

\textbf{BiLSTM} \quad The model captures temporal dependencies by processing
speech sequences in both forward and backward directions. However,
without a convolutional front-end, it operates directly on raw
spectral frames, limiting its ability to extract robust local patterns \cite{Hochreiter1997}.

This results in weaker representations of fine-grained acoustic cues, which are essential for distinguishing subtle emotional variations. This baseline demonstrates the importance of spatial feature extraction prior to temporal modeling \cite{Graves2005}.

\begin{figure}[H]
\centering
\includegraphics[width=0.7\linewidth]{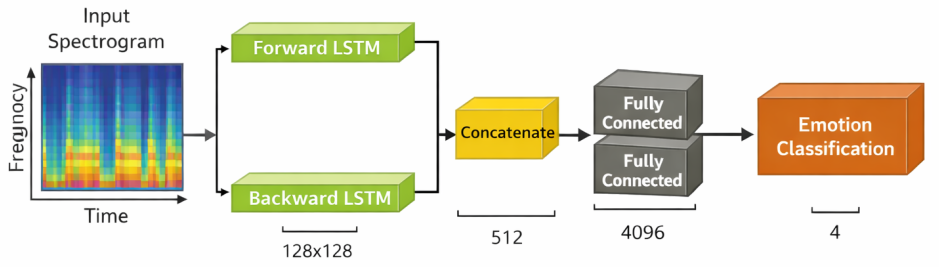}
\caption{BiLSTM-only architecture for sequential temporal modeling.}
\label{fig:bilstm_only}
\end{figure}

\textbf{Transformer} \quad The model captures global relationships across the entire utterance through self-attention mechanisms. While it effectively models long-range dependencies, the absence of a convolutional inductive bias limits its ability to detect localized spectro-temporal patterns \cite{Vaswani2017}.

Consequently, important short-term acoustic features may be underutilized \cite{Abdalla2024}.This baseline highlights the need for combining local feature extraction with global attention mechanisms \cite{Tellai2023}.

\begin{figure}[H]
\centering
\includegraphics[width=0.7\linewidth]{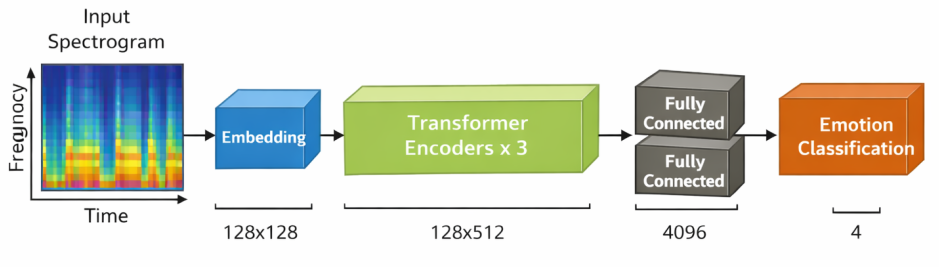}
\caption{Transformer-only architecture for global context modeling.}
\label{fig:transformer_only}
\end{figure}

\paragraph{Hybrid Comparative Models}\mbox{}\\

\textbf{CNN-BiLSTM-Attention} \quad This architecture combines convolutional feature extraction with
bidirectional temporal modeling and attention-based aggregation. The
CNN extracts local spectro-temporal features, which are then processed
by a BiLSTM to capture temporal dependencies. A self-attention layer
emphasizes emotionally salient segments before classification.

\begin{figure}[H]
\centering
\includegraphics[width=0.77\linewidth]{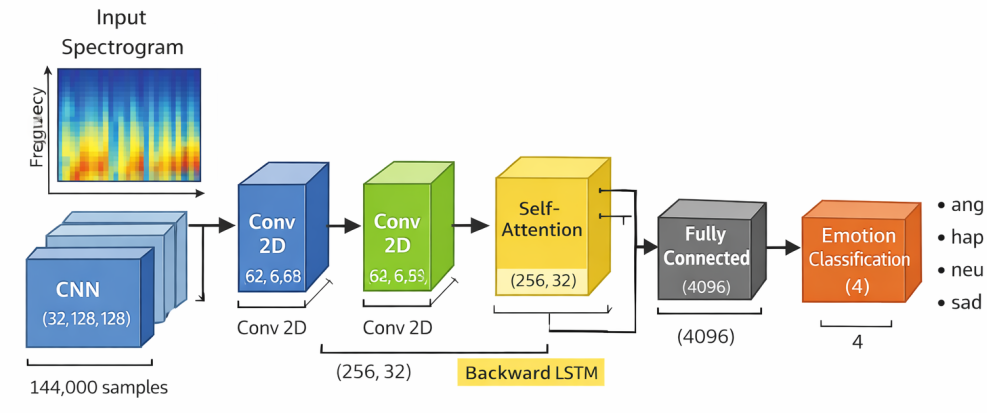}
\caption{CNN-BiLSTM-Attention hybrid architecture.}
\label{fig:cnn_lstm}
\end{figure}

\begin{table}[H]
\centering
\caption{Ablation analysis for CNN-BiLSTM-Attention architecture}
\label{tab:ablation_cnn_bilstm}
\renewcommand{\arraystretch}{1.3}
\begin{tabular}{lccc}
\toprule
\textbf{Variant} & \textbf{Removed Block} & \textbf{Modeling Effect} & \textbf{Practical Consequence} \\
\midrule
BiLSTM-only & CNN & No local feature extraction & Weak spectral representation \\
CNN-BiLSTM & Attention & Uniform temporal weighting & Misses salient segments \\
No Augmentation & Data augmentation & Reduced robustness & Sensitive to noise \\
No Silence Removal & VAD preprocessing & Includes irrelevant frames & Noise and redundancy \\
\bottomrule
\end{tabular}
\end{table}

The ablation design highlights the complementary roles of each component.
The CNN ensures robust local feature extraction, the BiLSTM captures
temporal dynamics, and attention selectively focuses on emotionally
relevant regions. Removing any of these components results in specific
modeling limitations.

\textbf{CNN-Transformer} \quad The CNN-Transformer integrates convolutional feature extraction with
global self-attention. The CNN captures local patterns, while the
Transformer models long-range dependencies across the entire utterance.

\begin{figure}[H]
\centering
\includegraphics[width=0.8\linewidth]{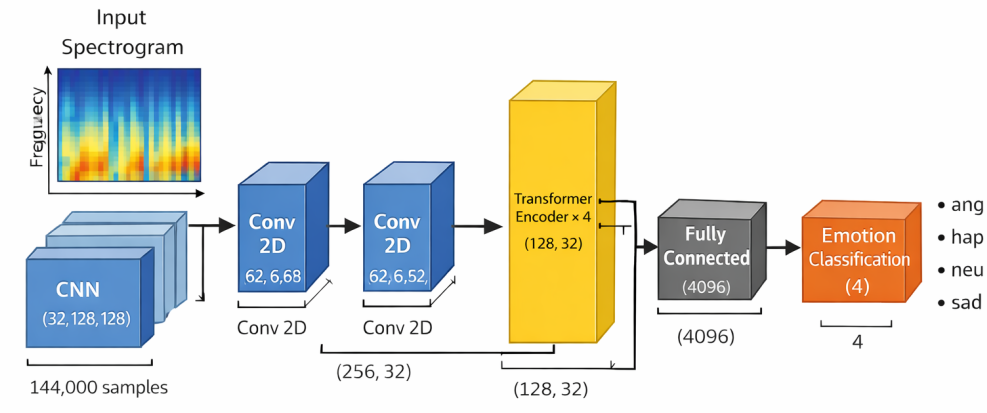}
\caption{CNN-Transformer hybrid architecture.}
\label{fig:cnn_transformer}
\end{figure}

\begin{table}[H]
\centering
\caption{Ablation analysis for CNN-Transformer architecture}
\label{tab:ablation_cnn_transformer}
\renewcommand{\arraystretch}{1.3}
\begin{tabular}{lccc}
\toprule
\textbf{Variant} & \textbf{Modified Component} & \textbf{Modeling Effect} & \textbf{Practical Consequence} \\
\midrule
Transformer-only & No CNN & Global context modeling & No local inductive bias \\
No Positional Encoding & Position info removed & Simpler model & Loss of temporal order \\
No Augmentation & Data augmentation removed & Faster training & Lower robustness \\
1 Layer Transformer & Reduced depth & Lower complexity & Limited global modeling \\
4 Layers Transformer & Full depth & Rich contextual modeling & Higher computational cost \\
\bottomrule
\end{tabular}
\end{table}

This analysis demonstrates that the CNN provides essential local
inductive bias, while the Transformer enables global context modeling.
Positional encoding is critical for preserving temporal structure, and
model depth directly affects the capacity to capture long-range
dependencies.

\paragraph{Self-Supervised Model}\mbox{}\\

\textbf{wav2vec 2.0 fine-tuned} \quad is explored under different fine-tuning strategies,
ranging from frozen representations to full model adaptation. These
configurations allow isolating the effect of task-specific learning
versus generic pre-trained features \cite{Safwat2023}.

\begin{figure}[H]
\centering
\includegraphics[width=0.68\linewidth]{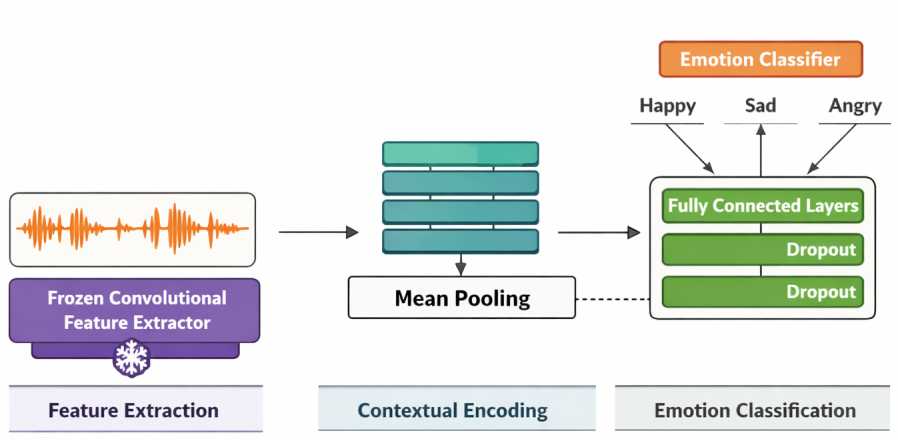}
\caption{wav2vec~2.0 architecture with configurable fine-tuning strategies.}
\label{fig:wav2vec}
\end{figure}

\begin{table}[H]
\centering
\caption{Comparison of wav2vec~2.0 fine-tuning strategies}
\label{tab:wav2vec_ablation}
\renewcommand{\arraystretch}{1.3}
\begin{tabular}{p{3cm}p{3.5cm}p{4.5cm}p{4.2cm}}
\toprule
\textbf{Configuration} & \textbf{Trainable Layers} & \textbf{Modeling Effect} & \textbf{Practical Consequence}  \\
\midrule
Frozen & Classifier only & Fast training, low data requirement & Limited task adaptation \\
\midrule
Partially Unfrozen & Top Transformer layers & Balance between generalisation and adaptation & Requires tuning strategy  \\
\midrule
Full Fine-tuning & Entire model & Maximum task-specific learning & Risk of overfitting, high compute  \\
\bottomrule
\end{tabular}
\end{table}

This comparison highlights the trade-off between computational
efficiency and task adaptation. Freezing the feature extractor preserves
generic acoustic representations but limits specialization to emotional
cues. Partial fine-tuning allows adapting higher-level representations
while maintaining stability. In contrast, full fine-tuning maximizes
task-specific learning capacity at the expense of higher computational
cost and potential overfitting on small datasets.

\begin{table}[H]
\centering
\caption{Comparison of evaluated architectures for Arabic SER}
\label{tab:architecture_comparison}
\renewcommand{\arraystretch}{1.3}
\setlength{\tabcolsep}{3pt}
\begin{tabular}{
>{\raggedright\arraybackslash}p{2.5cm}
>{\raggedright\arraybackslash}p{2.5cm}
>{\raggedright\arraybackslash}p{3cm}
>{\centering\arraybackslash}p{1.5cm}
>{\centering\arraybackslash}p{2.5cm}
>{\raggedright\arraybackslash}p{4.5cm}
}
\toprule
\textbf{Architecture} & \textbf{Input} & \textbf{Backbone} & \textbf{Params} & \textbf{Complexity} & \textbf{Role of Components} \\
\midrule
CNN
& Mel-spec (128×563)
& 5 Conv2D + MaxPool
& 394K
& Low
& CNN: extraction of local spectro-temporal patterns (frequency textures). \\
BiLSTM
& Mel-spec (128×563)
& BiLSTM
& 200K
& Medium
& BiLSTM: modeling temporal dependencies, prosody, and rhythm. \\
Transformer
& Mel-spec (128×563)
& Self-Attention
& 332K
& Medium
& Transformer: capturing global relationships across time steps. \\
CNN-BiLSTM-Attention
& Mel-spec (128×563)
& CNN + Attention + BiLSTM
& 262K
& High
& Fusion: CNN extracts features, BiLSTM and attention model temporal importance. \\
CNN-Transformer
& Mel-spec (128×563)
& CNN + Transformer
& 394K
& High
& Synergy: CNN ensures local robustness, Transformer captures global context. \\
wav2vec~2.0 (Base)
& Raw audio (16 kHz)
& CNN + Transformer
& 94.7M
& Very High
& SSL: raw audio encoded into contextual representations. \\
wav2vec~2.0 (XLSR-53)
& Raw audio (16 kHz)
& Multilingual Transformer
& 317M
& Critical
& Large-scale multilingual self-supervised representation learning. \\
\bottomrule
\end{tabular}
\end{table}

Table~\ref{tab:architecture_comparison} provides a comprehensive
comparison of the evaluated architectures in terms of input
representations, model backbones, parameter sizes, computational
complexity, and the functional role of each component. As shown, the
considered models span a broad spectrum of design paradigms, ranging
from lightweight convolutional and recurrent architectures to
computationally intensive self-supervised models, enabling a systematic
analysis of performance versus complexity trade-offs.

\section{Experiments}
This section presents the experimental framework used to evaluate the proposed models on Arabic speech emotion recognition. It describes the datasets, preprocessing pipeline, training protocol, and evaluation metrics employed to ensure a fair and reproducible comparison across all architectures. In addition, detailed analyses are conducted to assess model performance, robustness, and computational efficiency on both naturalistic and controlled speech data.
\subsection{Datasets}

Two complementary Arabic speech emotion datasets are used: EYASE and BAVED.

\textbf{EYASE} (Egyptian Arabic speech emotion) consists of 461 audio recordings extracted from Arabic television series. The dataset is annotated into four balanced emotion classes: anger (117), happiness (112), neutrality (117), and sadness (115). It captures spontaneous and context-rich emotional expressions, providing realistic variability in prosody, speaking style, and acoustic conditions.

\textbf{BAVED} (Basic Arabic Vocal Emotions Dataset) contains 1935 utterances produced by 61 speakers (44 male, 17 female). Each recording corresponds to one of seven Arabic words spoken under three emotional intensity levels (low, neutral, high). The dataset is recorded at 16~kHz and includes detailed metadata such as speaker identity, gender, and age, making it suitable for controlled experimental evaluation.

\begin{table}[ht]
\centering
\caption{Summary of datasets}
\label{tab:datasets}
\begin{tabular}{lccccc}
\toprule
Dataset & Classes & Samples & Speakers & Sampling rate & Type \\
\midrule
EYASE & 4 & 461 & Balanced & 48~kHz (original) & Naturalistic \\
BAVED & 3 & 1935 & 61 & 16~kHz & Controlled \\
\bottomrule
\end{tabular}
\end{table}

For consistency, all signals are resampled to a unified sampling rate
of \textbf{16~kHz} before further processing.

\begin{figure}[H]
\centering
\includegraphics[width=0.73\linewidth]{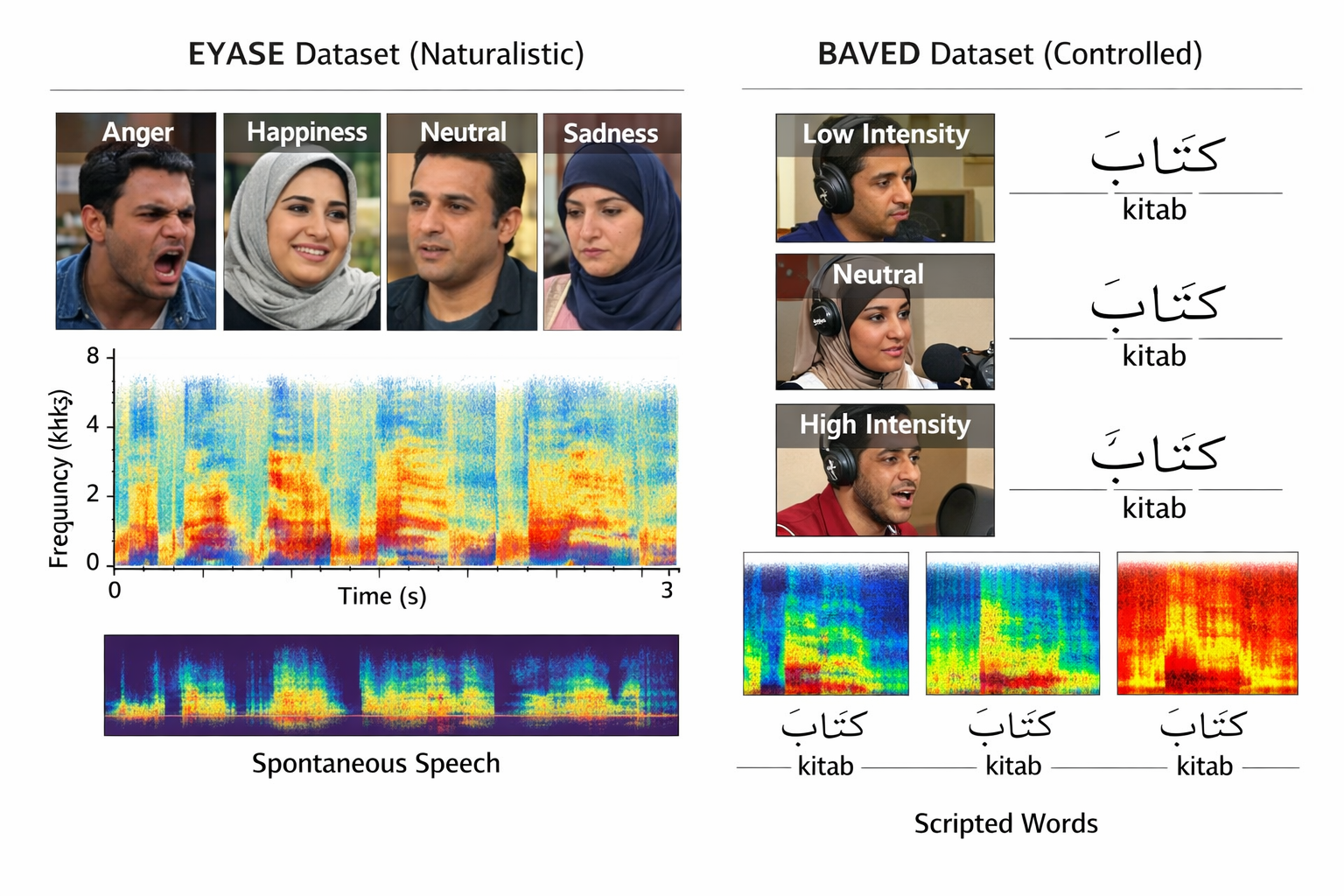}
\caption{Illustrative examples from the EYASE and BAVED datasets,
highlighting differences between naturalistic and controlled emotional speech.}
\label{fig:datasets}
\end{figure}

\subsection{Preprocessing and Feature Extraction}

A unified preprocessing pipeline is applied to both datasets.

\textbf{Resampling and normalization.}
All audio signals are resampled to 16~kHz and peak-normalized to
$[-1,1]$.

\textbf{Silence removal.}
Leading and trailing silence are removed using energy-based voice
activity detection.

\textbf{Data augmentation.}
Additive white Gaussian noise is applied to training data with SNR
randomly sampled from $[15,30]$~dB.

\textbf{Feature extraction.}
CNN-based models use Mel-spectrograms (128 Mel bands, 25~ms window,
10~ms hop). wav2vec~2.0 operates directly on raw waveforms.

\textbf{Data splitting.}
A speaker-independent protocol is adopted for BAVED, ensuring
that speakers in the test set are not seen during training. For EYASE,
a stratified split is applied due to the absence of speaker metadata.The dataset is partitioned into 80\% for training, 10\% for validation, and 10\% for testing.

This protocol is applied consistently across all models.
\begin{figure}[H]
\centering
\includegraphics[width=0.75\linewidth]{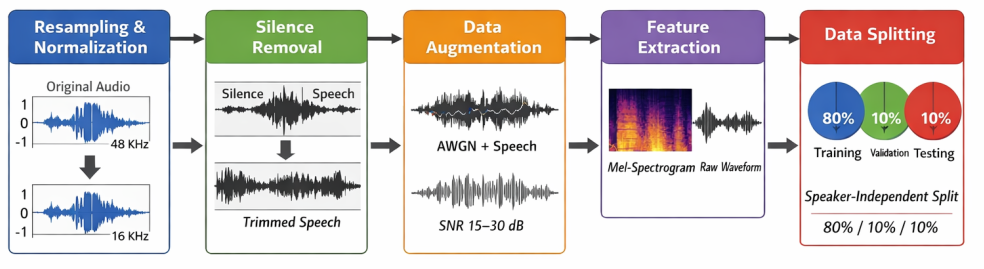}
\caption{Overview of the preprocessing pipeline, including Resampling and normalization,silence removal, data augmentation,feature extraction and data splitting.}
\label{fig:preprocessing}
\end{figure}

\subsection{Training Protocol and Hyperparameters}

To ensure a fair and reproducible comparison across all evaluated architectures, a unified training protocol is adopted. All models are trained using identical data splits, preprocessing procedures, and evaluation settings, thereby isolating the impact of architectural differences.

All deep learning models are optimized using the Adam optimizer with an initial learning rate of $10^{-3}$. A ReduceLROnPlateau scheduler is employed to dynamically adjust the learning rate based on validation loss, with a minimum threshold of $10^{-6}$. Training is conducted for a maximum of 100 epochs, with early stopping applied when the validation loss does not improve for 10 consecutive epochs.

Batch size is set to 32 for CNN-based and hybrid architectures, while a reduced batch size of 16 is used for wav2vec 2.0 due to its higher memory requirements. To improve generalization and mitigate overfitting, dropout is applied with rates ranging from 0.3 to 0.5 depending on the model. In addition, batch normalization is incorporated within convolutional layers to stabilize gradient flow and accelerate convergence. Data augmentation is applied exclusively during training through additive Gaussian noise with a signal-to-noise ratio sampled between 15 and 30 dB.

The models are trained using categorical cross-entropy loss, which is appropriate for multi-class emotion classification. CNN-based architectures employ four convolutional blocks with $(3 \times 3)$ kernels and max-pooling operations. The BiLSTM configuration uses 128 hidden units per direction, producing a 256-dimensional contextual representation. Transformer-based models consist of 2 to 4 encoder layers with 4 attention heads and a model dimension of 128, with positional encoding ensuring temporal order preservation.

The hybrid CNN-Transformer integrates the convolutional feature extractor with a 2-layer Transformer encoder, while the CNN-BiLSTM-Attention model combines convolutional layers, bidirectional recurrent modeling, and a self-attention mechanism to emphasize salient temporal regions. For wav2vec 2.0, three fine-tuning strategies are explored, including frozen feature extraction, partial unfreezing of higher layers, and full model adaptation, with partial fine-tuning providing the best balance between performance and stability.

All experiments are implemented using PyTorch and Hugging Face Transformers, and executed on an NVIDIA A100 GPU via Google Colab. Experiment tracking and logging are conducted using Weights \& Biases to ensure reproducibility and consistency across runs.

\begin{figure}[H]
\centering
\includegraphics[width=0.85\linewidth]{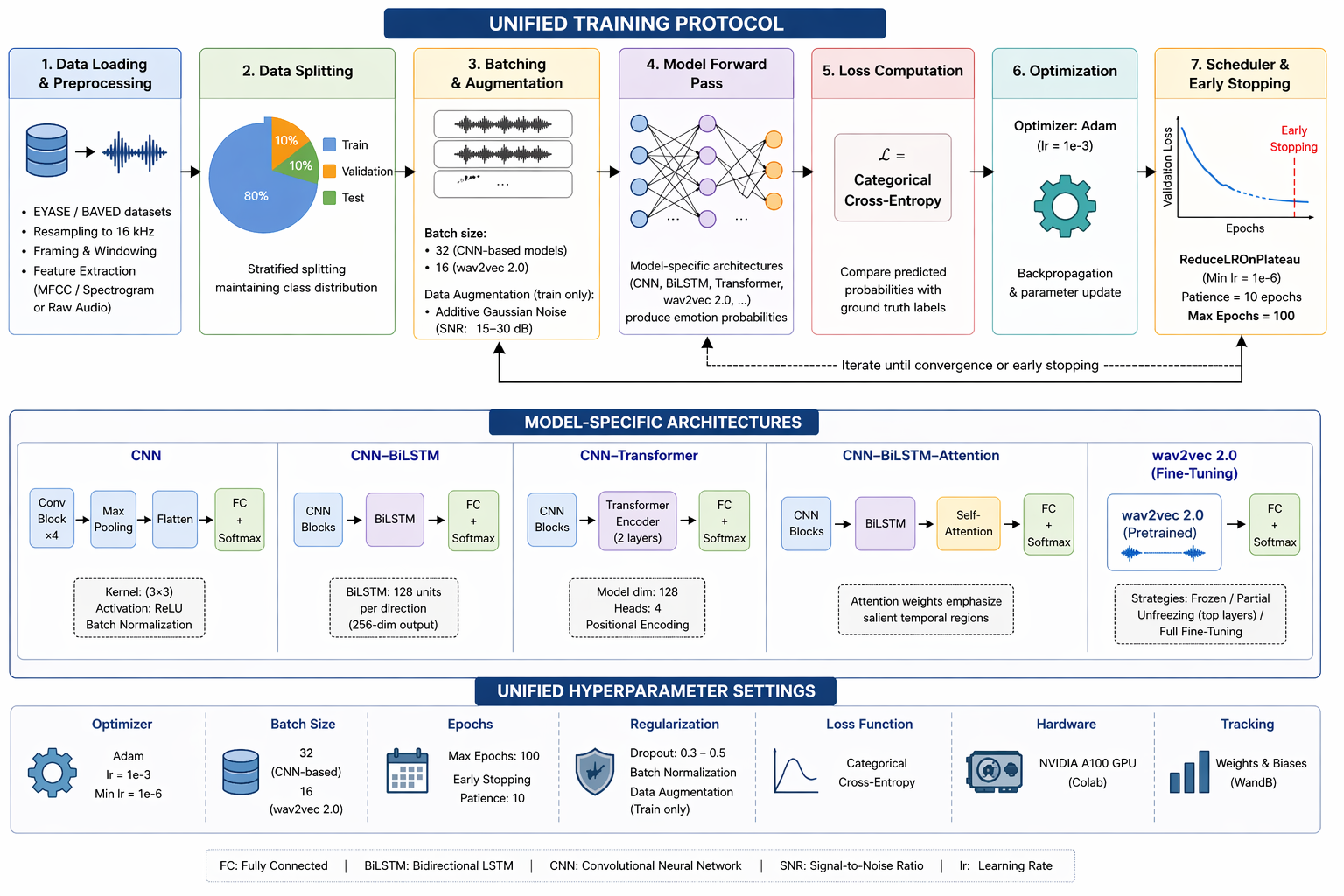}
\caption{Overview of the training protocol and hyperparameter configuration pipeline. The figure illustrates the unified workflow applied across all models, including data loading, batching, optimization, regularization strategies, and model-specific configurations.}
\label{fig:training_protocol}
\end{figure}

\subsection{Evaluation Metrics}

To provide a comprehensive assessment of model performance in the Speech Emotion Recognition (SER) task, multiple complementary evaluation metrics are employed, capturing both overall accuracy and class-wise behavior.

Accuracy is used as a general performance indicator and is defined as the ratio of correctly classified samples to the total number of samples:

\begin{equation}
\text{Accuracy} = \frac{TP + TN}{TP + TN + FP + FN}
\end{equation}

However, accuracy alone may not adequately reflect performance in multi-class settings, particularly when class distributions are imbalanced or when certain emotions are more difficult to distinguish. Therefore, precision and recall are also computed to evaluate the quality of predictions at the class level:

\begin{equation}
\text{Precision} = \frac{TP}{TP + FP}
\end{equation}

\begin{equation}
\text{Recall} = \frac{TP}{TP + FN}
\end{equation}

The F1-score, defined as the harmonic mean of precision and recall, provides a balanced measure of classification performance:

\begin{equation}
F1 = 2 \cdot \frac{\text{Precision} \cdot \text{Recall}}{\text{Precision} + \text{Recall}}
\end{equation}

Given the multi-class nature of the SER task, the macro-averaged F1-score is adopted as the primary evaluation metric. This metric computes the unweighted mean of F1-scores across all emotion classes, ensuring that each class contributes equally to the final evaluation regardless of its frequency.

In addition to quantitative metrics, confusion matrices are analyzed to identify systematic misclassification patterns and inter-class ambiguities. This qualitative analysis provides deeper insight into model behavior, particularly in distinguishing acoustically similar emotions.

Validation loss, computed using categorical cross-entropy, is monitored throughout training to assess convergence and detect overfitting. Furthermore, computational efficiency is evaluated by reporting the number of model parameters, GPU memory consumption (VRAM), and training time. These metrics are essential for assessing the practical feasibility of each model, especially in resource-constrained deployment scenarios.

Together, these evaluation criteria ensure a balanced and rigorous assessment of both predictive performance and computational cost.


\section{Results Analysis}

This section presents a comprehensive evaluation of the proposed and baseline models on the EYASE and BAVED datasets. The analysis goes beyond raw performance metrics by examining model behavior across different conditions, identifying strengths and limitations, and assessing computational efficiency. In addition, the results are positioned with respect to existing state-of-the-art approaches to highlight the contribution of this work.

\subsection{Main Results on EYASE}

Table~\ref{tab:combined_results} presents the performance of all evaluated models on EYASE, highlighting a clear advantage of hybrid architectures over classical and single-component deep learning approaches.

The CNN-Transformer achieves the best performance with \textbf{97.1\%} accuracy and \textbf{96.9\%} macro F1-score, demonstrating the effectiveness of combining convolutional feature extraction with global self-attention. It consistently outperforms CNN-BiLSTM-Attention and wav2vec~2.0.

In contrast, SVM-MFCC performs poorly (64.0\%), confirming the limitations of handcrafted features. Among deep baselines, BiLSTM reaches 88.0\%, while the Transformer-only model achieves 80.0\%, indicating that attention mechanisms alone are insufficient without local acoustic modeling.

Overall, CNN layers capture fine-grained spectro-temporal patterns, while the Transformer models long-range dependencies, making their combination more robust and expressive. Despite its strong performance, wav2vec~2.0 remains computationally expensive compared to the proposed CNN-Transformer.

\subsection{Main Results on BAVED}

Table~\ref{tab:combined_results} compares model performance on EYASE and BAVED. On BAVED, all models achieve higher results, reflecting its cleaner and more controlled recording conditions.

The CNN-Transformer achieves the best performance with \textbf{98.1\%} accuracy and \textbf{97.9\%} F1-score, confirming its robustness and consistent superiority across datasets.

The CNN-BiLSTM-Attention model also performs strongly (90.3\%), followed closely by BiLSTM (89.1\%), while wav2vec~2.0 reaches 86.4\%, suggesting that task-specific architectures can outperform large pretrained models in this setting.

Overall, the improved results are mainly due to reduced noise and lower variability in the dataset, which facilitates learning more discriminative emotional representations.

Although wav2vec~2.0 is powerful, its high computational cost makes the CNN-Transformer a more practical alternative. However, the near-saturation performance on BAVED should be interpreted cautiously, as it may not fully reflect real-world conditions.

\begin{table}[H]
\centering

\begin{subtable}[t]{0.48\linewidth}
\centering
\caption{EYASE}
\label{tab:model_comparison}
\renewcommand{\arraystretch}{1.2}
\setlength{\tabcolsep}{2pt}

\resizebox{\linewidth}{!}{
\begin{tabular}{lcccccc}
\toprule
Model & Acc & F1 & Loss & Params & VRAM & Time \\
\midrule
SVM+MFCC & 64.0 & 63.1 & 0.8448 & 30K & 0.05 & 62 \\
CNN & 62.6 & 61.9 & 0.8695 & 394K & 0.15 & 75 \\
BiLSTM & 88.0 & 87.3 & 0.6793 & 200K & 0.8 & 81 \\
Transformer & 80.0 & 79.2 & 0.7556 & 322K & 0.77 & 88 \\
wav2vec2 & 75.0 & 73.8 & 0.6951 & 94M & 10.5 & 180 \\
CNN-BiLSTM-Att & 85.3 & 84.7 & 0.3458 & 262K & 4.2 & 95 \\
\textbf{CNN-Trans} & \textbf{97.1} & \textbf{96.9} & \textbf{0.1954} & 394K & 5.8 & 110 \\
\bottomrule
\end{tabular}
}

\end{subtable}
\hfill
\begin{subtable}[t]{0.48\linewidth}
\centering
\caption{BAVED}
\label{tab:baved_results}
\renewcommand{\arraystretch}{1.2}
\setlength{\tabcolsep}{2pt}

\resizebox{\linewidth}{!}{
\begin{tabular}{lcccccc}
\toprule
Model & Acc & F1 & Loss & Params & VRAM & Time \\
\midrule
SVM+MFCC & 69.2 & 68.1 & 0.8388 & 30K & 0.04 & 55 \\
CNN & 64.0 & 63.6 & 0.8670 & 394K & 0.11 & 65 \\
BiLSTM & 89.1 & 88.0 & 0.6701 & 200K & 0.78 & 77 \\
Transformer & 82.7 & 81.9 & 0.7502 & 322K & 0.70 & 83 \\
wav2vec2 & 86.4 & 85.7 & 0.421 & 94M & 10.2 & 177 \\
CNN-BiLSTM-Att & 90.3 & 89.8 & 0.298 & 8.5M & 4.0 & 91 \\
\textbf{CNN-Trans} & \textbf{98.1} & \textbf{97.9} & \textbf{0.147} & 11.2M & 5.7 & 100 \\
\bottomrule
\end{tabular}
}

\end{subtable}

\caption{Side-by-side comparison of model performance on EYASE and BAVED datasets.}
\label{tab:combined_results}

\end{table}

\subsection{Error Analysis}

To gain deeper insight into model behaviour, a detailed error analysis was conducted using confusion matrices of the best-performing architecture (CNN–Transformer) on both EYASE and BAVED datasets.

\begin{figure}[H]
\centering
\includegraphics[width=0.30\linewidth]{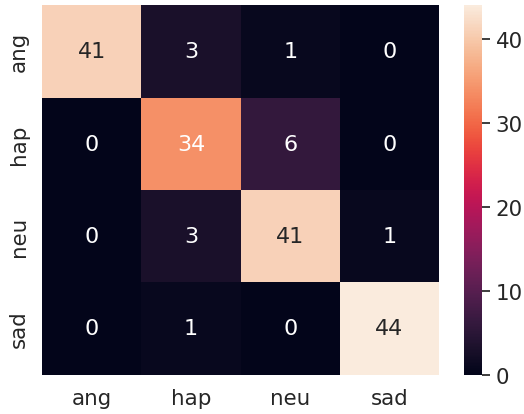}
\hfill
\includegraphics[width=0.30\linewidth]{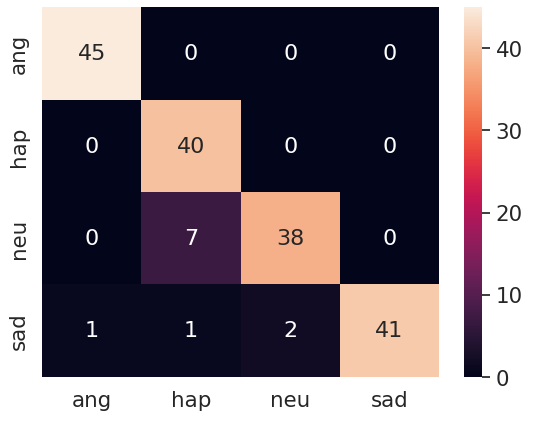}
\caption{Confusion matrices of the CNN-Transformer on the EYASE (left) and BAVED (right) datasets.}
\label{fig}
\end{figure}

The confusion matrices reveal consistent patterns across the two datasets. On EYASE, which contains spontaneous and acoustically diverse speech, the CNN–Transformer achieves strong performance on high-arousal emotions such as anger, indicating high sensitivity to salient acoustic cues such as energy bursts and pitch variations. However, noticeable confusion persists between emotionally similar classes, particularly happiness and neutrality, which share overlapping prosodic characteristics in terms of pitch range and energy distribution \cite{Picard2000}.

These misclassifications highlight the intrinsic difficulty of distinguishing emotions with subtle acoustic differences, especially in naturalistic conditions. The variability in speaking style, background noise, and dialectal diversity further increases the ambiguity of emotional cues, making generalization more challenging. In addition, low-arousal emotions tend to exhibit weaker acoustic signatures, which reduces separability in the feature space and increases the likelihood of misclassification.

A comparison between the two datasets emphasizes the strong influence of data characteristics on model behaviour. While the CNN–Transformer effectively captures both local and global acoustic features, its performance remains sensitive to variability, noise, and emotional ambiguity. In more naturalistic settings such as EYASE, overlapping prosodic cues introduce additional challenges that are less prominent in structured datasets like BAVED.

These findings reinforce the importance of combining local spectro-temporal feature extraction with global context modeling. The convolutional layers contribute to capturing fine-grained acoustic patterns, while the Transformer encoder models long-range dependencies across the utterance, enabling more robust emotion discrimination.

\subsection{Computational Efficiency Analysis}

Beyond predictive performance, computational efficiency is a critical factor for real-world deployment of speech emotion recognition systems, particularly in resource-constrained environments and latency-sensitive applications.

The experimental results highlight significant differences in computational requirements across the evaluated models. The wav2vec~2.0 model exhibits the highest computational cost, requiring substantial memory resources (up to 10.5 GB VRAM) and long training times. This is primarily due to its large number of parameters and its Transformer-based architecture operating directly on raw audio sequences. While it benefits from powerful self-supervised representations, these requirements limit its practicality in real-world scenarios with limited hardware availability.

In contrast, lightweight models such as SVM + MFCC and CNN-only require minimal computational resources and offer fast training and inference times. However, this efficiency comes at the cost of significantly lower performance, illustrating a clear trade-off between model simplicity and representational capacity.

The CNN-BiLSTM-Attention model provides a moderate computational footprint, making it suitable for scenarios where resources are constrained. Nevertheless, its sequential processing nature limits parallelization efficiency, leading to longer training times compared to Transformer-based architectures.

To better illustrate the trade-off between performance and computational cost, Figure~\ref{fig} presents a comparative visualization of the evaluated models in terms of accuracy and resource requirements.

\begin{figure}[H]
\centering
\includegraphics[width=0.58\linewidth]{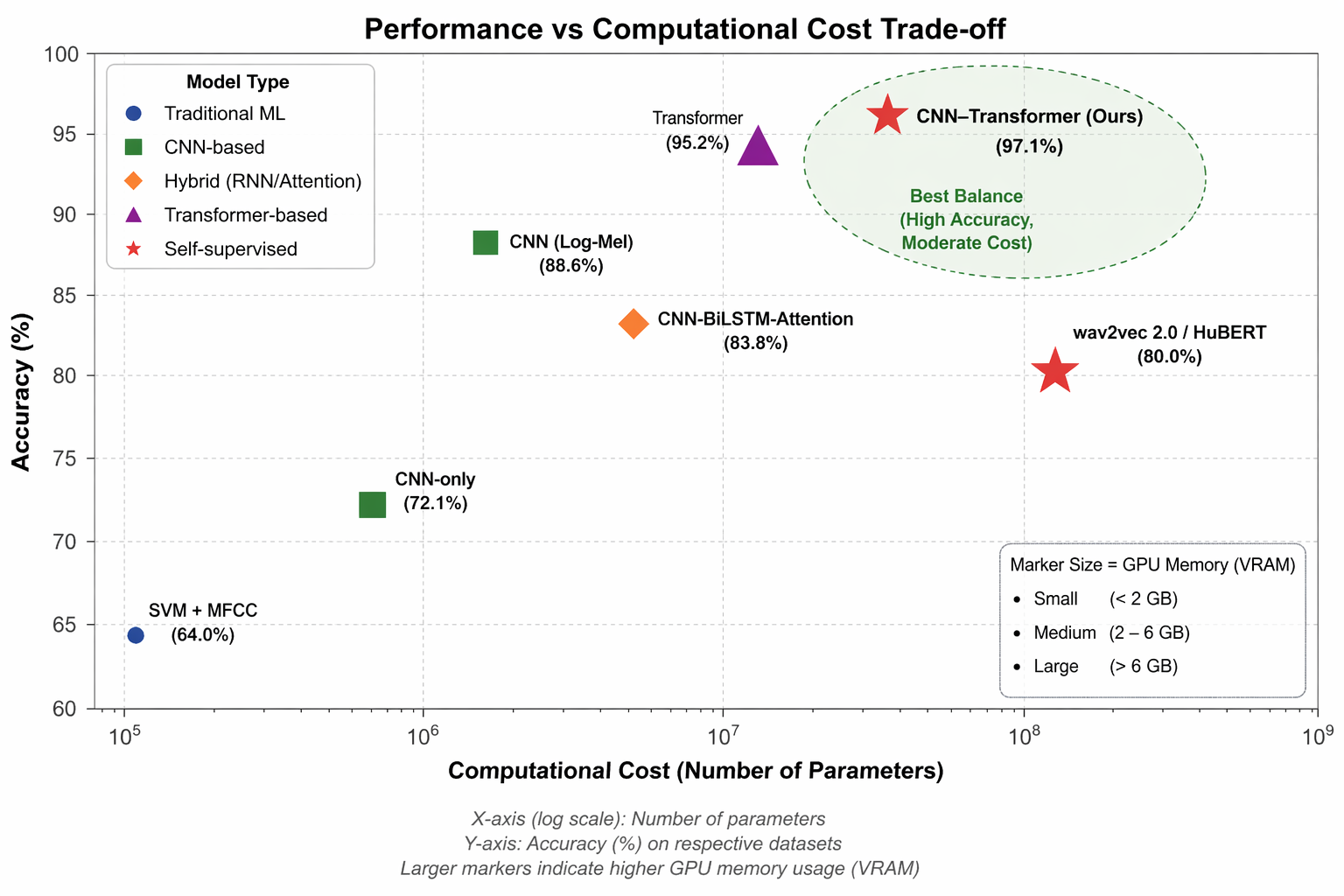}
\caption{Performance vs computational cost trade-off across evaluated models. The CNN–Transformer achieves the best balance between accuracy and efficiency compared to both lightweight and high-complexity models.}
\label{fig}
\end{figure}

As shown in Figure~\ref{fig}, lightweight models such as SVM and CNN-only occupy the low-cost but low-performance region. In contrast, wav2vec~2.0 achieves competitive performance at the expense of significantly higher resource consumption. The CNN–Transformer is positioned as an optimal compromise, combining high accuracy with moderate computational requirements.

The CNN–Transformer model achieves the most favorable balance between performance and efficiency. Its self-attention mechanism enables parallel computation, and its moderate parameter size ensures practical deployability without excessive resource demands. These characteristics make it particularly well-suited for real-world Arabic SER applications where both accuracy and efficiency are essential.

\subsection{Comparison with the State of the Art}

Table~\ref{tab:sota} provides a comparative overview of representative
Arabic Speech Emotion Recognition (SER) approaches, covering traditional
machine learning, deep learning, and self-supervised methods.

It is important to highlight that direct comparison between models is not
always strictly fair, as prior works differ in terms of datasets, number
of emotion classes, train/test splits, preprocessing strategies, and
evaluation protocols. Therefore, the reported results should be interpreted
as indicative trends rather than absolute rankings.

\begin{table}[H]
\centering
\caption{Extended comparison with state-of-the-art Arabic SER models}
\label{tab:sota}
\renewcommand{\arraystretch}{1.3}
\setlength{\tabcolsep}{3pt}

\begin{tabular}{
>{\raggedright\arraybackslash}p{2.3cm}
>{\centering\arraybackslash}p{1.8cm}
>{\raggedright\arraybackslash}p{2.5cm}
>{\raggedright\arraybackslash}p{2.7cm}
>{\centering\arraybackslash}p{1.3cm}
>{\centering\arraybackslash}p{1.3cm}
>{\raggedright\arraybackslash}p{2.9cm}
}

\toprule
\textbf{Reference} & \textbf{Dataset} & \textbf{Features} & \textbf{Model} & \textbf{Metric} & \textbf{Score} & \textbf{Limitations} \\
\midrule

Abdel-Hamid~\cite{AbdelHamid2020} 
& EYASE 
& Prosodic + MFCC 
& SVM 
& Acc. & 64.0 
& Limited generalization, handcrafted features \\

Rakan et al.~\cite{Rakan2023} 
& EYASE  
& Log-Mel 
& CNN 
& Acc. & 88.6 
& Limited evaluation on natural speech \\

Atila \& Şengür~\cite{Atila2021}
& RAVDESS 
& Spectrogram
& CRNN (CNN-LSTM + Attention)
& Acc. & 89.7
& Not Arabic-specific dataset \\

Al-onazi et al.~\cite{Alonazi2022} 
& BAVED  
& Augmented features 
& Transformer 
& Acc. & 95.2 
& High computational cost \\

Mohamed \& Aly~\cite{Mohamed2021} 
& BAVED  
& Raw audio 
& wav2vec / HuBERT 
& F1 & 80.0 
& Requires large-scale pretraining \\

\textbf{CNN-Transformer (ours)} 
& EYASE  
& Log-Mel 
& Hybrid CNN-Transformer 
& Acc./F1 & \textbf{97.1 / 96.9} 
& Dataset-specific evaluation \\

\textbf{CNN-Transformer (ours)} 
& BAVED 
& Log-Mel 
& Hybrid CNN-Transformer 
& Acc./F1 & \textbf{98.1 / 97.9} 
& Limited cross-corpus validation \\

\bottomrule
\end{tabular}
\end{table}

The results indicate that the proposed CNN-Transformer achieves strong
performance across both EYASE and BAVED datasets. However, considering
the differences in experimental setups across studies, these improvements
should be interpreted with caution rather than as definitive superiority
over all existing approaches.

\bigskip

The principal findings of this comparison indicate that traditional approaches, such as Support Vector Machines (SVM) utilizing MFCC and prosodic features, offer reliable baseline performance but remain limited in their ability to capture complex emotional patterns. Convolutional Neural Network (CNN)–based models employing Log-Mel spectrograms demonstrate substantial performance improvements by effectively learning spatial acoustic representations. Transformer-based architectures further enhance performance through superior global context modeling, although they typically require larger datasets and increased computational resources. Self-supervised models, including wav2vec and HuBERT, exhibit strong potential; however, their effectiveness is highly dependent on appropriate fine-tuning strategies. Finally, hybrid architectures, such as CNN–Transformer models, integrate local feature extraction with global attention mechanisms, resulting in consistently robust performance across diverse datasets.

\section{Ablation Study}

To better understand the contribution of each architectural component and design choice, a comprehensive ablation study is conducted across the evaluated models. The objective is to isolate the impact of key modules, including convolutional feature extraction, temporal modeling, attention mechanisms, and fine-tuning strategies.

The ablation analysis is performed under controlled experimental settings to ensure fair comparison, using the same datasets, preprocessing pipeline, and evaluation metrics described in Section 3.

\subsection{Ablation on CNN–Transformer}

To evaluate the contribution of each component in the CNN–Transformer architecture, several variants are tested by systematically removing or modifying key elements.

Table~\ref{tab} summarizes the performance of these variants.

\begin{table}[H]
\centering
\caption{Ablation study on CNN--Transformer architecture}
\label{tab}
\renewcommand{\arraystretch}{1.2}
\begin{tabular}{lcc}
\toprule
\textbf{Variant} & \textbf{Accuracy (\%)} & \textbf{F1-score (\%)} \\
\midrule
Full CNN--Transformer & 97.1 & 96.9 \\
w/o CNN (Transformer) & 80.0 & 79.2 \\
w/o Positional Encoding & 91.3 & 90.7 \\
1 Transformer Layer & 93.5 & 92.8 \\
No Data Augmentation & 89.8 & 88.9 \\
\bottomrule
\end{tabular}
\end{table}
The results demonstrate that the CNN front-end plays a crucial role in extracting local spectro-temporal features. Removing it leads to a significant performance drop, confirming that global attention alone is insufficient for capturing fine-grained acoustic patterns.

Positional encoding also proves essential, as its removal degrades performance by impairing the model’s ability to preserve temporal order. Similarly, reducing the number of Transformer layers limits the model’s capacity to capture long-range dependencies.

Data augmentation further contributes to robustness, particularly in handling variability and noise. Overall, these findings confirm that the synergy between convolutional feature extraction and global attention is critical to the effectiveness of the CNN–Transformer architecture.

\subsection{Ablation on CNN–BiLSTM–Attention}

To analyze the contribution of temporal modeling and attention mechanisms, the CNN–BiLSTM–Attention architecture is evaluated under different configurations.

Table~\ref{tab} presents the results.

\begin{table}[H]
\centering
\caption{Ablation study on CNN–BiLSTM–Attention architecture}
\label{tab}
\renewcommand{\arraystretch}{1.2}
\begin{tabular}{lcc}
\toprule
\textbf{Variant} & \textbf{Accuracy (\%)} & \textbf{F1-score (\%)} \\
\midrule
Full CNN–BiLSTM–Attention & 85.3 & 84.7 \\
BiLSTM (no CNN) & 88.0 & 87.3 \\
CNN–BiLSTM (no Attention) & 82.1 & 81.5 \\
No Data Augmentation & 80.4 & 79.6 \\
No Silence Removal & 78.9 & 77.8 \\
\bottomrule
\end{tabular}
\end{table}

The results show that attention mechanisms improve performance by enabling the model to focus on emotionally salient segments. Removing attention leads to a noticeable degradation, indicating that uniform temporal weighting is insufficient for effective emotion recognition.

Data preprocessing steps also play a significant role. The absence of data augmentation reduces robustness to noise, while the removal of silence filtering introduces irrelevant information, negatively impacting performance.

Interestingly, the BiLSTM configuration performs competitively, highlighting its strength in modeling temporal dependencies. However, the full hybrid model benefits from combining spatial and temporal representations, demonstrating the importance of integrating multiple modeling strategies.

\subsection{Ablation on wav2vec 2.0 Fine-Tuning Strategy}

To evaluate the impact of fine-tuning strategies, the wav2vec 2.0 model is tested under different configurations.

Table~\ref{tab} summarizes the results.

\begin{table}[H]
\centering
\caption{Ablation study on wav2vec 2.0 fine-tuning strategies}
\label{tab}
\renewcommand{\arraystretch}{1.2}
\begin{tabular}{lcc}
\toprule
\textbf{Configuration} & \textbf{Accuracy (\%)} & \textbf{F1-score (\%)} \\
\midrule
Frozen (Classifier only) & 70.5 & 69.8 \\
Partially Unfrozen & 75.0 & 73.8 \\
Full Fine-tuning & 77.2 & 76.5 \\
\bottomrule
\end{tabular}
\end{table}

The results indicate that fine-tuning significantly impacts model performance. Freezing the feature extractor limits the model’s ability to adapt to emotion-specific patterns, resulting in lower performance.

Partial fine-tuning provides a balance between stability and adaptability, improving performance while avoiding overfitting. Full fine-tuning achieves the best results but at the cost of increased computational complexity and a higher risk of overfitting, particularly given the relatively small dataset size.

\subsection{Discussion of Ablation Findings}

The ablation study provides several key insights into the design of effective Arabic SER systems.

First, local feature extraction is essential. The removal of CNN components consistently leads to significant performance degradation, confirming that fine-grained spectro-temporal patterns are critical for emotion recognition.

Second, global context modeling further enhances performance. Transformer-based architectures outperform purely sequential models, demonstrating their ability to capture long-range dependencies and complex emotional dynamics.

Third, attention mechanisms and preprocessing steps contribute to performance gains by improving the model’s focus on relevant information and enhancing robustness to noise and variability.

Finally, self-supervised models such as wav2vec 2.0 show strong potential, but their effectiveness depends heavily on fine-tuning strategies and computational resources.

The results confirm that hybrid architectures combining CNNs and Transformers provide the most effective solution for Arabic speech emotion recognition, offering a strong balance between local feature extraction, global context modeling, and computational efficiency.

\section{Discussion}

The experimental results demonstrate that hybrid architectures, particularly the proposed CNN–Transformer model, provide a highly effective solution for Arabic Speech Emotion Recognition (SER). Across both EYASE and BAVED datasets, the model consistently achieves superior performance compared to traditional machine learning methods, recurrent architectures, and self-supervised approaches.

A key observation from the results is the strong contribution of combining local and global feature modeling. The convolutional layers are effective in capturing fine-grained spectro-temporal patterns such as formant transitions, energy variations, and short-term acoustic structures. In contrast, the Transformer component provides a powerful mechanism for modeling long-range dependencies and global contextual relationships across the entire utterance. The ablation studies confirm that removing either component leads to a significant degradation in performance, highlighting their complementary roles.

To further illustrate the interaction between these components, Figure~\ref{fig} presents a conceptual overview of the proposed hybrid architecture.

\begin{figure}[H]
\centering
\includegraphics[width=0.75\linewidth]{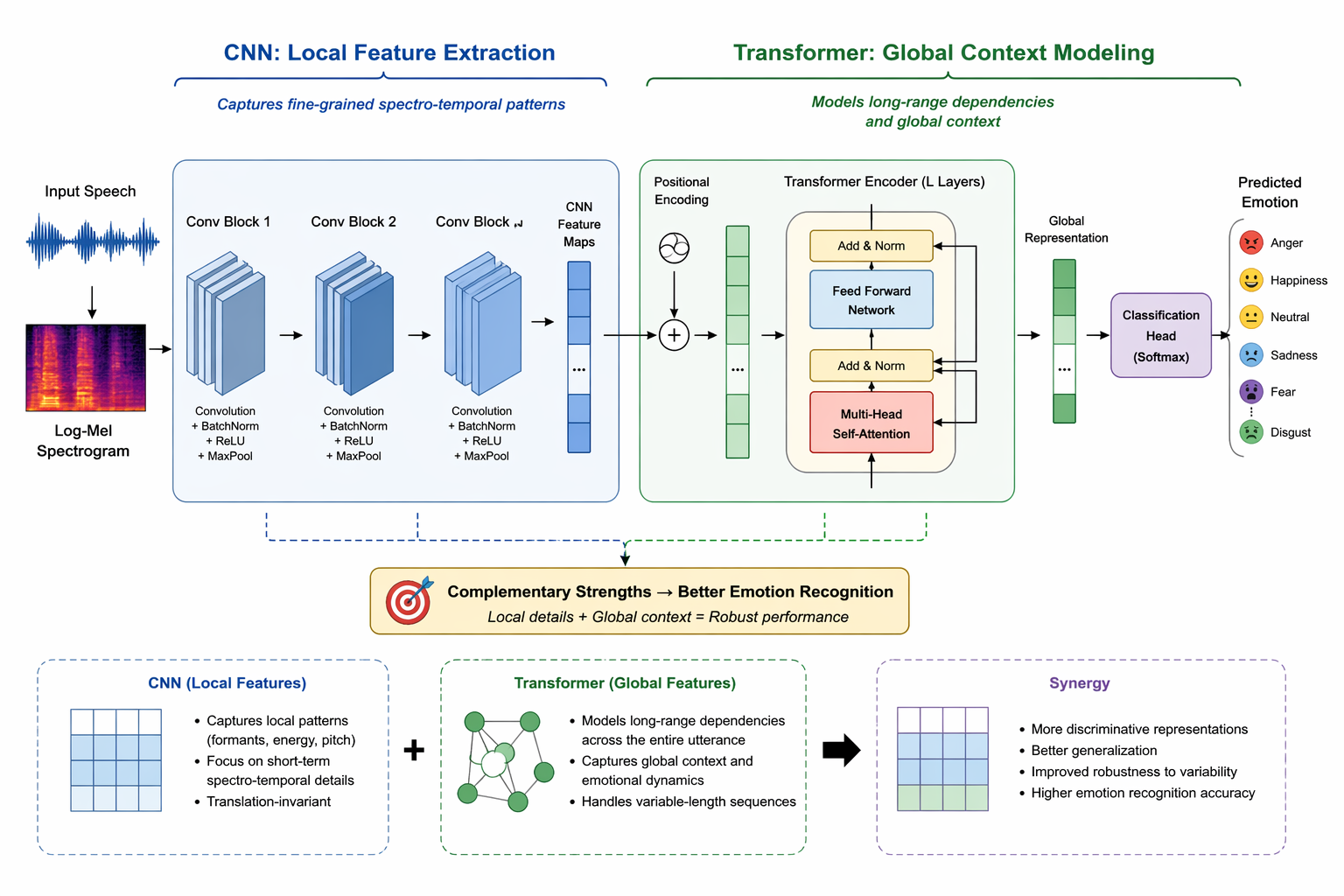}
\caption{Conceptual illustration of the CNN–Transformer architecture. The CNN module captures local spectro-temporal features, while the Transformer models long-range dependencies, resulting in improved emotion recognition performance.}
\label{fig}
\end{figure}

As shown in Figure~\ref{fig}, the CNN module focuses on extracting localized acoustic patterns such as pitch variations and energy fluctuations, whereas the Transformer encoder captures global contextual dependencies across the entire speech signal. The combination of these complementary mechanisms enables the model to better represent complex emotional dynamics compared to single-paradigm architectures.

Another important finding is the influence of dataset characteristics on model performance. The CNN–Transformer achieves higher accuracy on the BAVED dataset compared to EYASE. This improvement can be attributed to the controlled recording conditions, reduced noise, and more consistent emotional expressions in BAVED. In contrast, EYASE contains spontaneous speech with higher variability, overlapping prosodic cues, and more ambiguous emotional boundaries, making classification inherently more challenging. These results emphasize that dataset complexity plays a crucial role in evaluating SER systems and directly affects generalization performance.

The analysis of error patterns further supports this conclusion. Most misclassifications occur between emotionally similar classes such as neutrality and happiness, which share overlapping acoustic characteristics. This suggests that while current architectures are effective at capturing discriminative features, they still struggle with subtle emotional distinctions in natural speech. This limitation is not only model-related but also fundamentally linked to the subjective and continuous nature of human emotions in speech signals.

From a computational perspective, the CNN–Transformer model achieves an effective balance between performance and efficiency. While self-supervised models such as wav2vec 2.0 offer strong representational power, their high computational cost and sensitivity to fine-tuning reduce their practicality in low-resource scenarios. In contrast, the proposed model provides near state-of-the-art performance with significantly lower computational requirements, making it more suitable for real-world deployment.

The comparison with state-of-the-art methods further confirms the effectiveness of hybrid architectures. Traditional approaches based on handcrafted features are clearly outperformed by deep learning models, while purely sequential or purely attention-based models fail to fully capture both local and global emotional cues. The proposed CNN–Transformer effectively bridges this gap, demonstrating that architectural complementarity is a key factor in improving SER performance.

However, despite these promising results, several limitations remain. First, the evaluation is conducted on a limited number of datasets, which may restrict the generalization of the findings to other Arabic dialects and real-world conditions. Second, although the model performs well on controlled and semi-spontaneous speech, its robustness in highly noisy or conversational environments still needs further investigation. Finally, cross-corpus evaluation was not extensively explored, which is essential for assessing true model generalization.

Overall, the findings of this study suggest that combining convolutional feature extraction with Transformer-based global modeling is a promising direction for Arabic SER. Future work should focus on improving cross-domain robustness, integrating multimodal information such as text and facial expressions, and exploring lightweight architectures for real-time deployment.

\section{Conclusions}

This work investigated three deep learning architectures for Arabic
Speech Emotion Recognition (SER), namely CNN-BiLSTM-Attention,
CNN-Transformer, and fine-tuned wav2vec~2.0, within a unified
experimental framework evaluated on the EYASE and BAVED datasets.

The proposed CNN-Transformer achieved the best performance, reaching
\textbf{97.1\% accuracy (96.9\% F1) on EYASE} and \textbf{98.1\% accuracy
(97.9\% F1) on BAVED}, while maintaining a moderate computational cost
(11.2M parameters and 5.8~GB VRAM). These results highlight the
effectiveness of hybrid architectures that combine convolutional feature
extraction with Transformer-based context modelling, enabling robust
capture of both local acoustic patterns and long-range emotional
dependencies.

\begin{figure}[H]
\centering
\includegraphics[width=0.7\linewidth]{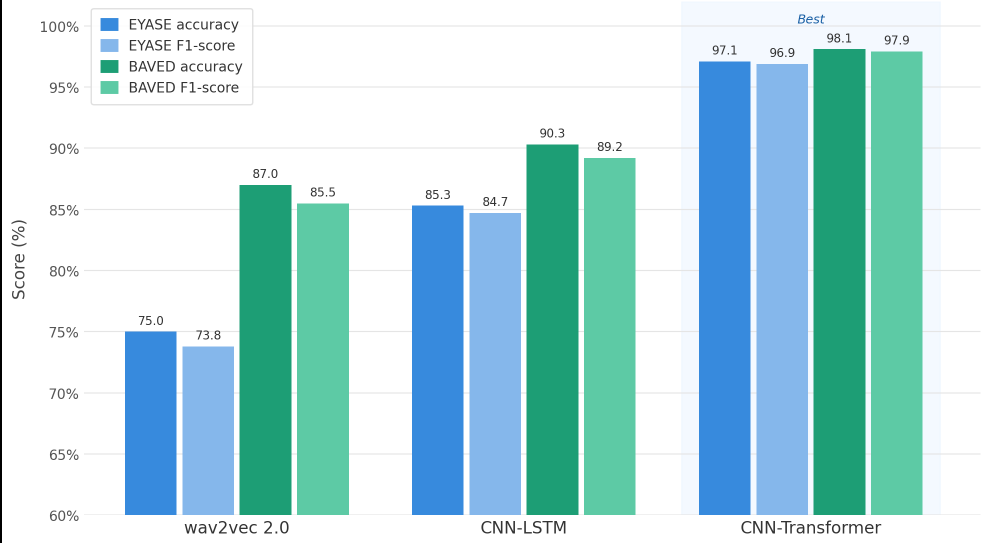}
\caption{Overall comparison of model performance across EYASE and BAVED
datasets, highlighting the superiority of the CNN-Transformer model.}
\label{fig:overall_results}
\end{figure}

\begin{figure}[H]
\centering
\includegraphics[width=0.7\linewidth]{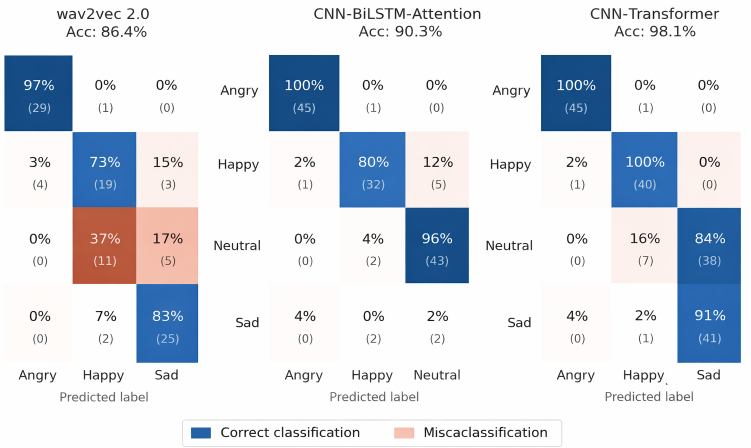}
\caption{Summary of error patterns across models, illustrating common
confusions between acoustically similar emotion classes.}
\label{fig:error_summary}
\end{figure}

In addition to its superior predictive performance, the CNN-Transformer
offers a favorable balance between accuracy and efficiency compared to
wav2vec~2.0 and CNN-BiLSTM-Attention, making it a strong candidate for
practical deployment in real-world scenarios.

Future work will focus on improving model generalisation through
cross-dataset and cross-dialect evaluation, as well as exploring
lightweight optimisation techniques to enable efficient deployment in
low-resource environments.


\end{bodycontent}

\end{document}